\documentclass[12pt,a4paper]{article} 

\pdfoutput=1
\usepackage[numbers]{natbib}
\usepackage[affil-it]{authblk}

\usepackage{amsmath}
\usepackage{amsthm}
\usepackage[english]{babel}
\usepackage{graphicx}
\usepackage{amsfonts} 
\usepackage{amssymb} 
\usepackage{hyperref} 
\usepackage{float}
\usepackage{times}
\usepackage{tensor}
\usepackage{mathtools}
\usepackage{color}
\usepackage{bbold}

\usepackage{subdepth}

\newcommand{\pd}[1]{\frac{\del}{\del #1}}
\newcommand{\PD}[2]{\frac{\del #1}{\del #2}}

\newcommand{\ket}[1] {\ensuremath {\left| #1 \right\rangle }}

\newcommand{\phat}[0]{\hat{p}}
\newcommand{\qhat}[0]{\hat{q}}
\newcommand{\del}[0]{\partial}

\title{Quantum mechanics on York slices}
\author{Philipp Roser\thanks{proser@clemson.edu}}
\affil{\footnotesize Department of Physics and Astronomy,\\ Clemson University, Kinard Laboratory,\\ Clemson, SC 29631-0978, USA}
\date{\vspace{-0.5cm}}		

\begin{document}
 \maketitle
\begin{abstract}
For some time the York time parameter has been identified as a candidate for a physically meaningful time in cosmology. An associated Hamiltonian may be found by solving the Hamiltonian constraint for the momentum conjugate to the York time variable, although an explicit solution can only be found in highly symmetric cases. The Poisson structure of the remaining variables is not canonical. Here we quantise this dynamics in an anisotropic minisuperspace model via a natural extension of canonical quantisation. The resulting quantum theory has no momentum representation. Instead the position basis takes a fundamental role. We illustrate how the quantum theory and the modified representation of its momentum operators lead to a consistent theory in the presence of the constraints that arose during the Hamiltonian reduction. We are able to solve for the eigenspectrum of the Hamiltonian. Finally we discuss how far the results of this model extend to the general non-homogeneous case, in particular perturbation theory with York time.
\end{abstract}

\section{Introduction}
The canonical quantisation of general relativity carries with it the notorious problem of time, one of whose facets is the existence of a Hamiltonian constraint in the classical theory, which manifests after quantisation in the form of the Wheeler-deWitt equation, leading to an apparently `frozen' dynamics for the universe \citep{deWitt1967}, \citep{Kuchar2011}, \citep{Rovelli2007}, \citep{Thiemann2008}, \citep{Kiefer2012}. The Hamiltonian constraint arises because of the time-reparame\-terisation invariance of the classical theory. One possibility to overcome this difficulty is to break this invariance by identifying a physically meaningful time parameter from among the spatial and/or matter variables. One particularly promising suggestion has been the `York parameter' $T$ \citep{ChoquetBruhatYork1980} --- roughly speaking, equal to the fractional rate of spatial contraction of the local volume element --- whose physical significance has been made apparent through its importance in the initial-value problem of general relativity \citep{York1971,York1972,York1973}. Two further independent motivations for its use as a fundamental time parameter are found in \citep{QadirWheeler1985} and \citep{Valentini1996}. The emerging time found in shape dynamics \citep{GomesGrybKoslowski2011, BarbourKoslowskiMercati2013ProbOfTime,Mercati2014} also bears a close relationship to York time.

Having chosen York time as a preferred temporal parameter, one can obtain a physical, non-vanishing Hamiltonian $H_{phys}$ by solving the Hamiltonian constraint for the momentum $P_T$ conjugate to the York parameter $T$, a procedure known as Hamiltonian reduction. However, in the general case the equation is too difficult to solve by known methods \citep{ChoquetBruhatYork1980}. In \citep{RoserValentini2014a} we performed this procedure for the case of a homogeneous and isotropic minisuperspace model with scalar fields and analysed the subsequent quantum theory. However, the simplicity of the model used there hid another aspect of this theory: the fact that the dynamical variables left over after the extraction of the York parameter and conjugate momentum, namely the scale-free components $\tilde{g}_{ij}=g^{-\frac13}g_{ij}$ of the metric and associated momentum $\tilde{\pi}^{ij}=g^\frac13(\pi^{ij}-\frac13 g^{ij}Tr(\pi))$ are not canonical. Their Poisson brackets are \citep{ChoquetBruhatYork1980}
\begin{align}
 \{\tilde{g}_{ab}(\vec{x}),\tilde{pi}^{cd}(\vec{y})\}   &= \left(\delta_a^{(c}\delta_b^{d)}-\tfrac13\tilde{g}_{ab}\tilde{g}^{cd}\right)\delta^3(\vec{x}-\vec{y}) \label{gpiPB}\\
 \{\tilde{\pi}^{ab}(\vec{x}),\tilde{\pi}^{cd}(\vec{y})\} &= \frac13\left(\tilde{g}^{cd}\tilde{\pi}^{ab}-\tilde{g}^{ab}\tilde{pi}^{cd}\right)\delta^3(\vec{x}-\vec{y}). \label{pipiPB}
\end{align}

In this paper we explore this Poisson structure and the resulting quantum theory by using another pure-gravity minisuperspace model which is homogeneous but anisotropic, and therefore displays the non-canonical Poisson brackets. In the classical theory the choice of York time may be seen as a mere gauge choice. However, since different choices of time parameter lead to the quantisation of different variables, different choices of time parameter can lead to physically distinct quantum theories. A cosmological constant is easily included in this model.

We show that a consistent quantum theory that incorporates all the constraints of the classical theory can indeed be formulated, and we solve the Hamiltonian eigenequation. An interesting aspect that emerges from the Poisson structure, in particular the non-commutativity of the momenta with each other, is the absence of a momentum representation.

We hope that by considering this analytically solvable model insights into the structure of the full quantum (not minisuperspace) theory may be gained. The full theory unfortunately remains unsolvable due to the aforementioned difficulty in solving the Hamiltonian constraint. However, the generalisation of the results to cosmological perturbation theory, where an approximate physical Hamiltonian can be derived, is discussed in section \ref{CommentsOnPertTheory}.

\section{Classical Theory}
In order to study the consequences of this Poisson structure it is sufficient to consider a homogeneous anisotropic vacuum minisuperspace model with a spatial metric consisting of diagonal components only, $g_{ij}=\delta_{ij}Q_i$. It is easy to see that the associated momenta must also be diagonal, $\pi^{ij}=\delta^{ij}P^i$. The classical solution of the Einstein equations with these restrictions are the so-called Kasner models \citep[sec.\ 30.2]{MisnerThorneWheeler1973}.

Because of the `compression' of two indices ($g_{ij},\pi^{ij}$) into one ($Q_i,P^i$) in our formulation of the dynamics, we employ the following summation convention: Indices are summed over if they appear at least once as an upper and a lower index each. Indices may appear multiple times as either upper or lower indices only however without implying a summation. All indices are spatial indices only and range of $1,2,3$. It is furthermore conventient to introduce the inverse metric variables, $q^a\equiv(q_a)^{-1}$.

The York parameter is defined to be proportional to the fractional rate of contraction of space, $T\equiv 2\pi/3\sqrt{g}$, with the constant of proportionality chosen such that its conjugate momentum is the negative of the local volume element, $P_T=-\sqrt{g}$, where $g\equiv \det(g_{ab})=Q_1Q_2Q_3$ is the metric determinant and $\pi\equiv Tr(\pi^{ab})=Q_iP^i$ is the trace of the momenta.\footnote{In general, the York parameter can only function as a viable choice of time if an appropriate slicing condition is satisfied, ensuring that it is indeed constant across each slice. However, here its constancy is automatically guaranteed by spatial homogeneity.} That is, the York parameter and momentum extract the isotropic `scale' component of the variables. The remaining scale-free variables are $\tilde{g}_{ij},\tilde{\pi}^{ij}$ as defined above for the general case, or
\begin{equation} q_a\equiv g^{-\frac13}Q_a, \qquad p^a\equiv g^\frac13(P^a-\tfrac13\pi Q^a).\end{equation}
in the case of the present model. In virtue of their definition they obey the constraints
\begin{align} q_1q_2q_3&=1, \label{scaleconstraint} \\ q_ap^a &=0. \label{tracefreeconstraint} \end{align}
The tracelessness \ref{tracefreeconstraint} of $p^a$ ensures that the first constraint \ref{scaleconstraint}, the scale-free condition, is preserved.

In terms of $q_a,p^a$ the Hamiltonian constraint, obtained by following the ADM procedure \citep{ADM1962}, takes the form
\begin{equation}\label{HamiltonianConstraint} 
  0=\mathcal{H}=2\kappa\left[ -\frac{\pi^2}{6\sqrt{g}}+\frac{1}{\sqrt{g}}q_a^{2}p^{a2}\right]=2\kappa\left[\tfrac38T^2P_T-\frac{1}{P_T}q_a^2p^{a2}\right],
\end{equation}
where $2\kappa\equiv16\pi G$. The physical Hamiltonian associated with York time is given by $H_{phys}=-P_T(q_a,p^a,T)$, where $P_T(q_a,p^a,T)$ is the function obtained when solving the Hamiltonian constraint for $P_T$ in terms of the other variables. In the full theory the analogous equation is a difficult elliptic equation with no known general solutions. Here however it is a simple quadratic, yielding
\begin{equation}\label{ClassicalHamiltonian} H_{phys}\equiv-P_T=\pm\left[\frac{8}{3T^2}q_a^2p^{a2}\right]^\frac12.\end{equation}
The choice of sign is not physical. For any given physical trajectory corresponding to one sign choice there is a corresponding solution for the other sign choice, characterised by $q_a(T)\rightarrow q_a(T)$, $p^a(T)\rightarrow-p^a(T)$. Since the physical interpretation of the numerical value of $H_{phys}$ is that of `volume' however and volume is conventionally defined as positive, we assume the positive sign in \ref{ClassicalHamiltonian}.

The original variables $Q_i,P^i$ are canonical, $\{Q_a,P^b\}=\delta_a^b$ with other Poisson brackets vanishing. The new variables on the other hand have a more complicated Poisson structure,
\begin{align}
 \{q_a,p^b\} &= \delta_a^k-\tfrac13q_aq^b, \label{qpPB}\\
 \{p^a,p^b\} &= \tfrac13\left(p^aq^b-p^bq^a\right) \label{ppPB},
\end{align}
obtained by using the definitions of $q_a$ and $p^a$ in terms of $Q_a$ and $P^a$ and the canonical Poisson brackets $\{Q_a,P^b\}$. It is straightforward to verify that the motion generated by the momenta, $q_a\rightarrow q_a+\epsilon_b\{q_a,p^b\}$, $p^a\rightarrow\epsilon_b\{p^a,p^b\}$ for a small vector $\epsilon_b$, preserves the constraints \ref{scaleconstraint}, \ref{tracefreeconstraint}. The same holds true for motion generated by the Hamiltonian, $q_a\rightarrow q_a+\delta T\{q_a,H\}$, $p^a\rightarrow p^a+\delta T\{p^a,H\}$, although the tracelessness constraint \ref{tracefreeconstraint} is required to show that the scale-free condition \ref{scaleconstraint} is preserved.

Specifically, the equation of motion are
\begin{align} 
 q_a^\prime  &= \{q_a,H\} = \sqrt{\frac{8}{3T^2\cdot q_k^2p^{k2}}}\cdot q_b^2p^b\delta^b_a \label{qeom}\\
 p^{a\prime} &= \{p^a,H\} = -\sqrt{\frac{8}{3T^2\cdot q_k^2p^{k2}}}\cdot q_bp^{b2}\delta^a_b, \label{peom}
\end{align}
where a prime ($^\prime$) denotes derivation with respect to York time $T$.

Solutions to these equations may be found by inspection using the fact that $p^{b\prime}=-q^bp^b\delta^{ab}q_a^\prime$. One finds solutions
\begin{equation} q_a(T)=(-4/3T)^{2(s_a-\frac13)}, \qquad p^a(T) = (s_a-\tfrac13)(-4/3T)^{-2(s_a-\frac13)}, \label{classicalsolutions} \end{equation}
with constant parameters $s_a$ satisfying $s_1+s_2+s_3=1$, $s_1^2+s_2^2+s_3^2=1$. These solutions are exactly the Kasner models. In order to see this and to get a better intuition of the relation of York time $T$ and standard cosmological time $t$ in these models, recall that in the Kasner models $g=t^2$ and the general fact that $T$ was defined as$-4/3$ times the fractional rate of change of volume, so that
\begin{equation} T= -\frac{4}{3t}. \end{equation}
This makes it apparent that \ref{classicalsolutions} are indeed the Kasner solutions. The value of $H_{phys}$ is given by $-P_T=\sqrt{g}$, so that $H_{phys}=t$ --- cosmological time is just the numerical value of the physical Hamiltonian.

The fact that one obtains exactly the same solutions illustrates the consistency of the reduced formalism.

A cosmological constant may be included in the above formalism, leading to the substitution $(8/3T^2)\rightarrow(\frac38T^2-2\Lambda)^{-1}$ in $H_{phys}$. The solutions of the equations of motion are then 
\begin{align}
 q_a(T) &= \gamma_a \left\vert T+\sqrt{T^2-\frac{16}{3}\Lambda}\right\vert^{+2(s_a-\frac13)} \label{qsolution}\\
 p^a(T) &= \gamma_a^{-1} (s_a-\tfrac13) \left\vert T+\sqrt{T^2-\frac{16}{3}\Lambda}\right\vert^{-2(s_a-\frac13)} \label{psolution}
\end{align}
where the parameters $s_a$ satisfy the same condition as in the Kasner model, $\gamma_a$ are constants chosen to satisfy eq.\ \ref{scaleconstraint} and $T$ is restricted to $\frac38T^2>2\Lambda$. For a discussion of the appropriate range of values of $T$ and the question of extending $T$ over the entire real line, see \citep{Roser2015CosmExtension}.

\section{Quantisation and representations}

The canonical quantisation recipe provides a list of instructions of turning a classical theory with canonical variables into a viable quantum theory. In our case however the Poisson brackets are not canonical and there is no such established recipe. However, there is a natural extension of that recipe that one may consider to provide a good guess for what may constitute a viable quantised theory.

As usual, we `promote' variables to operators, which are to act on a space of quantum states $\Psi$. Poisson brackets become commutator brackets and the right-hand side of \ref{qpPB} and \ref{ppPB} gain a factor $i\hbar$ on dimensional grounds to match the `missing' power of momentum as compared to the left-hand side. The proposed quantum theory is therefore defined by the commutator expressions
\begin{align}
 \label{qpcommutator} [\qhat_a,\phat^b] &= i\hbar\left(\delta_a^b-\tfrac13\qhat_a\qhat^b\right), \\
 \label{ppcommutator} [\phat^a,\phat^b] &= \frac{i\hbar}{3}\left[\phat^a\qhat^b-\phat^b\qhat^a\right]
\end{align}
and other commutators vanish. Constraints $\phi(q_i,p^j)=0$ act as operators on states as $\phi(\qhat_i,\phat^j)\Psi_{phys}=0$ if $\Psi_{phys}$ is to be considered a physically possible state.

One may now readily see why this theory does not allow for a momentum representation, that is, a representations of quantum states as functions $\tilde{\psi}(p^i)$ of state-space coordinates $p^i$ such that $\phat^a\tilde{\psi}(p^i)=p^a\tilde{\psi}(p^i)$. The existence of such a representation directly contradicts the non-commutativity of the momenta. However, there is a position representation satisfying the commutation relations \ref{qpcommutator}, \ref{ppcommutator}, given by
\begin{align}
 \qhat_a\Psi(q_i) &= q_a\Psi(q_i), \\
 \phat^b\Psi(q_i) &= \left[-i\hbar\left(\delta^b_a-\tfrac13q^bq_a\right)\pd{q_a}\right]\Psi(q_i).
\end{align}
We discuss the implications of the non-existence of a momentum representation in section \ref{Conclusion}.

We are left to consider the constraints. The tracelessness requirement in operator form, 
\begin{equation} \qhat_a\phat^a=0, \end{equation}
is satisfied identically in this representation, so that it does not restrict the space of physical wavefunctions. On the other hand, the absence of absolute scale,
\begin{equation} \qhat_1\qhat_2\qhat_3\Psi_{phys}=\mathbb{1} \Psi_{phys} \end{equation}
implies that physical wavefunctions must vanish off the two-dimensional surface $\Sigma$ given by $q_1q_2q_3=0$. On the full configuration space physical wavefunctions are therefore discontinuous at $\Sigma$. One may be worried that this is problematic given that one encounters expressions of the form $\partial/\partial q_i\;\Psi$, which are derivatives across the discontinuity. However, derivatives in the position representation only appear in the form of linear combinations as given by the momenta, which are directional derivatives tangent to the constraint surface. This is, of course, the quantum result corresponding to the fact that classically momenta generate motion within the constraint surface. Specifically, considering a basis for the tangent space to $\Sigma$ at an arbitrary point by writing $q_3=(q_1q_2)^{-1}$ and calculating the corresponding tangent vectors,
\begin{align}
 T_1 \equiv\left(\PD{q_1}{q_1},\PD{q_2}{q_1},\PD{q_3}{q_1}\right) &= \left(1,0, -1/q_1^2q_2\right), \\
 T_2 \equiv\left(\PD{q_1}{q_2},\PD{q_2}{q_2},\PD{q_3}{q_2}\right) &= \left(0,1, -1/q_1q_2^2\right),
\end{align}
the momenta can be expressed in terms of this basis,\footnote{Furthermore, one notes that the momenta are not linearly independent operators.}
\begin{align}
 \phat^1 &=\left(\tfrac23 T_1-\frac{q_2}{3q_1}T_2\right)\cdot\nabla \\
 \phat^2 &=\left(-\frac{q_1}{3q_2}T_1+\tfrac23 T_2\right)\cdot\nabla \\
 \phat^3 &=\left(-\tfrac13q_1^2q_2T_1-\tfrac13q_1q_2^2T_2\right)\cdot\nabla.
\end{align}
The evolution of the value of $\Psi$ at some point $\vec{q}$ is therefore `blind' to the properties of $\Psi$ outside the constraint surface.

The dynamics of the quantum theory is determined by the quantised, time-dependent Hamiltonian,
\begin{equation} \hat{H} = \sqrt{\frac{8}{3T^2}}\widehat{\sqrt{q_i^2p^{i2}}}. \label{HamiltonianOperator}\end{equation}
Since classically the numerical value of the Hamiltonian is `volume', in the quantum theory $\hat{H}$ defines a `volume spectrum' with volume eigenfunctions and eigenvalues rather than the more conventionally encountered energy spectrum \citep{RoserValentini2014a}.

The obvious difficulty with $\hat{H}$ is the appearance of a `square-root' operator, whose meaning is not initially clear. Another question is the factor-ordering ambiguity in the radicand. Regarding the latter, one can show using the commutation relations that changing the ordering of $(\qhat_a\qhat_a\phat^a\phat^a)$ either has no effect or merely corresponds to adding or subtracting a constant $\frac43\hbar^2$ or $\frac83\hbar^2$. A different ordering choice therefore corresponds to a shift in the Hamiltonian eigenvalues but does not change the eigenfunctions. Where necessary, we will therefore with minimal loss of generality assume the ordering `$qqpp$', also expressible in the form
\begin{equation} \qhat_a\qhat_a\phat^a\phat^a\Psi(q)=\left(q_a^2\del^{a2}-\tfrac13q_aq_b\del^a\del^b+\tfrac23q_a\del^a\right)\Psi(q). \end{equation}

The square-root operator can be dealt with formally by assuming the existence of a series expansion,
\begin{equation} \hat{H}=(8/3T^2)^\frac12\sum_{n=0}^\infty w_{n,a}(q)(\del^a)^n,\end{equation}
which acts straightforwardly on a state when expressed as a Fourier transform,\footnote{Note however that the Fourier expansion does not constitute a integral over momentum eigenstates since no momentum representation exists. In fact, the relation of the Fourier components $k^c$ and the momenta can be seen via the action of the latter on the former,
  \begin{equation} \phat^a\psi(q)=\hbar\int d^3k\; \tilde{\psi}(k)\cdot(\delta^a_b-\tfrac13q^aq_b)k^b\,e^{ik^cq_c}
				 =\hbar k^a\psi(q)-\frac{\hbar}{3}\int d^3k\; \tilde{\psi}(k)\cdot q^aq_b k^b\,e^{ik^cq_c}. 
  \end{equation} }
\begin{equation} \psi(q)=\int d^3k\; \tilde{\psi}(k)e^{ik^cq_c},\end{equation}
giving
\begin{align} \hat{H}\left(\tilde{\psi}(k)e^{ik^cq_c}\right)
 &= \int d^3k\;\sum_{n=0}^\infty w_{n,a}(q)\,(ik^a)^n\cdot\tilde{\psi}(k)e^{ik^cq_c} \notag\\
 &= \int d^3k\;\sqrt{q_a^2(ik^a)^2-\tfrac13q_aq_b(ik^a)(ik^b)+\tfrac23q_a(ik^a)}\;\cdot\tilde{\psi}(k)e^{ik^cq_c}. 
\end{align}
This formal result does not seem however very practical or insightful. 

Instead we wish to deal with the square-root operator explicitly. Since analysis of $\hat{H}^2$ is a lot easier than $\hat{H}$ itself (as there is no square root and so the interpretation of the operator expression is clear) it is useful to establish the relationship between operators $\hat{h}\equiv(\qhat_a^2\phat^{a2})^\frac12$ and $\hat{f}\equiv\hat{h}^2=\qhat_a^2\phat^{a2}$.

If $\ket{h}$ is an eigenfunction of $\hat{h}$ with eigenvalue $h$, $\hat{h}\ket{h}=h\ket{h}$, then it is also an eigenfunction of $\hat{h}^2$ with eigenvalue $h^2$, $\hat{h}^2\ket{h}=h^2\ket{h}$. The converse is in general not true: there may be eigenfunctions of $\hat{f}=\hat{h}^2$ which are not eigenfunctions of $\hat{h}$. However, if $\hat{f}$ is diagonalisable with a complete set of eigenstates (this will follow from Hermiticity established below), then these also form a set of eigenstates for $\hat{h}=\hat{f}^\frac12$ with square-rooted eigenvalues.

Regarding Hermiticity, if $\hat{h}$ is Hermitian, then so is $\hat{h}^2$. Again, for arbitrary operators the converse is not true: If $\hat{A}^2=\hat{B}$ and $\hat{B}$ is Hermitian, then it is not guaranteed that $\hat{A}$ is Hermitian. However, if $\hat{A}=\hat{B}^\frac12$ is defined in terms of the diagonalised operator, then $\hat{A}$ will be Hermitian too (up to a subtlety also established below for the case of $\hat{f}$: $\hat{B}$ must be positive semi-definite, that is, it must have only non-negative eigenvalues, so that their square roots are real). That is, not every square root of a Hermitian operator is Hermitian, but one can always find one via diagonalisation. This is how we intend to interpret $\hat{h}=(\qhat_a^2\phat^{a2})^\frac12$ in terms of $\hat{f}$.

The operator $\hat{f}$ is Hermitian only on $\Sigma$, not on the full, unconstrained configuration space. In order to establish its Hermiticity it is useful to perform a change of coordinates $(q_1,q_2,q_3)\rightarrow(u\equiv q_1,v\equiv q_2, w\equiv q_1q_2q_3)$, so that $w=1$ describes the constraint surface, facilitating the integration. With $\partial_u=\partial^1-(w/u^2v)\partial^3$, $\partial_v=\partial^2-(w/uv^2)\partial^3$, $\partial_w=(1/uv)\partial^3$, the momenta are
\begin{align}
 \phat^1 &=-i\hbar\left(\tfrac23\del_u-\frac{v}{3u}\del_v\right) \\
 \phat^2 &=-i\hbar\left(\tfrac23\del_v-\frac{u}{3v}\del_u\right) \\
 \phat^3 &= \frac{i\hbar}{3}\left(\frac{u^2v}{w}\del_u + \frac{uv^2}{w}\del_v\right),
\end{align}
the volume element is $dq_1dq_2dq_3=(uv)^{-1}dudvdw$ and the operator $\hat{f}$ is
\begin{equation}\hat{f}\equiv\qhat_a\qhat_a\phat^a\phat^a = -\frac{2\hbar^2}{3}\Big[u^2\del_u^2+v^2\del_v^2-uv\del_u\del_v+v\del_v+u\del_u\Big]. \label{finuv}\end{equation}
In terms of these coordinates the Hermiticity of $\hat{f}$ on the constraint surface,
\begin{equation}\int_\Sigma \psi^\dag (\hat{f}\chi) =\int_\Sigma (\hat{f}\psi)^\dag\chi,\end{equation}
is easily shown. Note, however, that the momenta themselves are not Hermitian and therefore do not constitute `observables' in the conventional sense. For example,
\begin{equation} \int\limits_\Sigma \psi^\dag(\phat^1\chi) 
    = \int_\Sigma (\phat^1\psi)^\dag\chi - \int_0^\infty\int_0^\infty du\;dv\;\frac{1}{uv}\cdot\frac{i\hbar}{u}\psi^\dag\chi. 
\end{equation}

One reason we chose to examine the simplest model in which the Poisson brackets are non-trivial is that it allows us to explore their quantisation without having to worry about too many other cumbersome notational or other details. Another is that the Hamiltonian eigenequation, that is, the `volume eigenspectrum' can be solved exactly. This is because $\hat{f}$ is a homogeneous operator (eq.\ \ref{finuv}) whose eigenfunctions may be readily found by inspection,
\begin{equation} \phi_{n,m}(u,v)=A_{n,m}u^nv^m, \end{equation}
with eigenvalues
\begin{equation} \hat{f}\phi_{n,m}(u,v) =-\frac{2\hbar^2}{3}\big[n^2+m^2-nm\big]\phi_{n,m}(u,v).\end{equation}
Na\"ively $m,n\in\mathbb{C}$, although the values will be restricted shortly. With the interpretation of the square-root operator discussed above, the Hamiltonian eigensolutions are
\begin{equation} \hat{H}\phi_{n,m}(u,v) = h_{n,m}(T)\;\phi_{n,m}(u,v),\qquad h_{n,m}(T)=i\;\frac{4\hbar}{3|T|}\sqrt{n^2+m^2-nm}.\end{equation}
The eigenfunctions $\phi_{m,n}(u,v)$ are, however, not normalisable on $u,v\in(0,\infty)$ for any values $n,m$. But they are bounded (and, in fact, of constant magnitude) for purely imaginary $n,m$, and divergent for all other values. That is, let $n=i\beta$, $m=i\delta$, $\beta,\delta\in\mathbb{R}$. Then
\begin{equation} h_{i\beta,i\delta}(T) = -\;\frac{4\hbar}{3|T|}\sqrt{\beta^2+\delta^2-\beta\delta}, \end{equation}
where $\beta^2+\delta^2-\beta\delta\geq0$ always, so that the eigenvalues are real. Thus the reality of eigenvalues is equivalent with the non-divergence of the eigenfunctions. This is closely analogous to the `plane wave' eigenfunctions of a free particle in basic particle quantum mechanics. There the eigenfunctions are of the form $\exp(ikx)$ for real $k$ and therefore not normalisable either but bounded with constant magnitude. Imaginary values of $k$ are excluded even though they solve the eigenvalue equation because they entail divergent eigenfunctions. The analogy can be made more apparent by writing $\phi_{i\beta,i\delta}=e^{i(\beta\ln u+\delta\ln v)}$. The eigenvalues are sign-definite, that is, either all positive or all negative, depending on the choice of sign in the Hamiltonian. As we did in the classical theory we can choose `volume' to be positive, so that the Hamiltonian is a positive-semidefinite operator. We see furthermore that there is a unique minimum-volume state, $\phi_{0,0}$, which has constant eigenvalue zero, somewhat analogous to a `vacuum' state.

One can also derive uncertainty relations. However, since the momenta are not observables the relevance of these relation is questionable and their meaning is obscure. Nonetheless we include them for completeness. One finds
\begin{equation} \sigma_{q_i}\sigma_{p^j} \geq \frac{\hbar}{3}\qquad (i=j) \end{equation}
but the other non-zero relations are not as well-behaved and state-dependent,
\begin{align}
 \sigma_{q_i}\sigma_{p^j} &= \infty, \qquad (i\neq j)\qquad \text{ for H-eigenstates}\\
 \sigma_{p^i}\sigma_{p^j} &= 0 \text{ or } \infty \text{ depending on symmetry properties of chosen state}.
\end{align}

Including a cosmological constant in the quantum theory is relatively straightforward since only the time-dependent pre-factor of the Hamiltonian changes, with the resulting Hamiltonian eigenvalues changing accordingly. The eigenfunctions remain the same.

\section{Application to the general and perturbation theory}\label{CommentsOnPertTheory}

For a physically more relevant analysis one must, of course, abandon the assumption of homogeneity and consider the general Poisson brackets \ref{gpiPB}, \ref{pipiPB}. As discussed above, the associated Hamiltonian constraint cannot be solved for $P_T$ as is required in order to derive the explicit functional form of the physical Hamiltonian associated with York time. While this ultimately presents a difficulty that would have to be overcome in order to derive the general non-linear theory, certain properties of the resulting quantum theory can already be described. In particular, analogously to the derivation above, there is no momentum basis but only a position basis (conformal superspace) where
\begin{align} \hat{g}_{ab}(x)\Psi[g_{ij}] &= g_{ab}(x)\Psi[g_{ij}], \\
		 \hat{\pi}^{ab}(x)\Psi[g_{ij}] &= -i\hbar\left(\delta_c^{(a}\delta_d^{b)}-\tfrac13g^{ab}g_{cd}\right)\frac{\delta}{\delta g_{cd}(x)}\Psi[g_{ij}].
\end{align}
Just as in the homogeneous model above the `tracelessness' constraint 
\begin{equation}\hat{g}_{ab}\hat{\pi}^{ab}\Psi_{phys}[g_{ij}]=0\end{equation} 
is satisfied identically in the quantum theory and imposes no restriction on the physicality of states. On the other hand, the `scalefree' constraint 
\begin{equation}\hat{g}\Psi_{phys}[g_{ij}]=\Psi_{phys}[g_{ij}]\end{equation}
implies that physical states have support only on the classical constraint surface given by $\det(g_{ij})=1$. Momenta once again act tangentially, allowing for a consistent formulation of the quantum theory in the position basis.

While the explicit Hamiltonian is not known for the general theory, it is possible to construct cosmological perturbation theory based on York time \cite{RoserInPreparation}. Here one considers a homogeneous background on which one imposes gravitational and matter perturbations, which are assumed small. This allows one to solve for the physical Hamiltonian based on the York parameter of the homogeneous background only. One then proceeds to exploit the gauge ambiguity (equivalently, the ambiguity of the original slicing at the perturbative level) of the perturbative degrees of freedom such that the originally chosen `background' York time is, in fact, the exact slicing. Thus one obtains a perturbative Hamiltonian, which in virtue of the perturbative expansion consists of a sum of quadratic terms, albeit with strongly time-dependent coefficients as well as mixed momentum-position terms. The dynamics of the gravitational part of the perturbations is then determined by the Poisson structure discussed above. 

In particular, suppose there is an approximately homogeneous background slicing and the exact `York gauge' has been chosen. Then suppose we expand the reduced variables as
\begin{equation} \tilde{g}_{ij}=\gamma_{ij}+h_{ij},\qquad\tilde{\pi}^{ij}=\tilde{\bar{\pi}}^{ij}+\nu^{ij}, \end{equation}
where $\gamma_{ij}$ and $\tilde{\bar{\pi}}^{ij}$ are the reduced `variables` of the background, which are respectively constant and vanishing. By substituting these expressions into the Poisson brackets \ref{gpiPB} and \ref{pipiPB} one readily obtains brackets for the perturbation variables,
\begin{align} 
\{h_{ab}(x),\nu^{cd}(y)\} 
     &= \Big[ \delta_a^{(c}\delta_b^{d)} - \tfrac13\gamma_{ab}\gamma^{cd} + \tfrac13\gamma^{cd}h_{ab}-\tfrac13\gamma_{ab}\gamma^{ce}h_{ef}\gamma^{fd} \hspace{0.125\textwidth} \notag \\
									&\hspace{0.33\textwidth}+\text{(2nd order terms)}\Big]\;\delta^3(x-y) \label{hnuPB} \\ 
\{\nu^{ij}(x),\nu^{kl}(y)\} &= \tfrac13\left[\gamma^{cd}\nu^{ab}-\gamma^{ab}\nu^{cd}+\text{(2nd order terms)}\right]\;\delta^3(x-y). \label{nunuPB}
\end{align}
Terms of all orders in the perturbation variables appear because of the expansion of the inverse reduced metric, $\tilde{g}^{ij}=\gamma^{ij}-\gamma^{ia}h_{ab}\gamma^{bj}+\gamma^{ia}h_{ab}\gamma^{bc}h_{cd}\gamma^{dj}-\dots$ only. Note that the momentum-momentum bracket \ref{nunuPB} has no zero-order terms, so that this bracket is `almost canonical' in a more concrete sense than that of \citep{ChoquetBruhatYork1980}. This also implies that there is an `approximate momentum representation'. The position-momentum bracket \ref{hnuPB} on the other hand is non-canonical even at zeroth order, although the terms are constant across all space and time in virtue of containing background variables only.

Further details of this procedure will be discussed in a future paper. Suffice to say, this does not, of course, constitute a fundamental theory but one may hope that the dynamics obtained in this manner resembles the perturbative limit of dynamics generated by the elusive solution to the Hamiltonian constraint of the complete, non-linear theory.

\section{Conclusion}\label{Conclusion}

In this paper we showed how a consistent quantum theory may be defined based on non-canonical brackets via an obvious extension of the canonical quantisation recipe. The model chosen to explore was complicated enough for the unusual Poisson structure to become apparent, yet simple enough to allow us to solve for the `volume spectrum' of the Hamiltonian.

One particular aspect of this Poisson and commutator bracket structure is the absence of a momentum basis. Unlike in theories with canonical variables, `position' and `momentum' are therefore not equally valid bases for the state space of the quantum theory. Contingent on the idea that the York parameter does, in fact, constitute a physically preferred time coordinate, this suggests that the position basis, that is, \emph{configuration space} is the natural arena in which to describe quantum physics. There may be an argument that this suggests taking more seriously `interpretations' of quantum theory that already consider the configuration-space description as physically preferred, such as the de~Broglie-Bohm `pilot-wave' formulation \citep{deBroglie1928,Bohm1952a,Bohm1952b,Bell1987, Holland1993}. However, this is of no concern here. Furthermore, we note that in any case the preferred position basis arises only with respect to the gravitational degrees of freedom, not those of matter, so one must be careful not to overstate the significance of this `preferred basis'. The fact that the momentum-momentum bracket is `almost canonical' in the case of perturbations is however not significant in this regard. While an approximate momentum representation may be found, it is merely a mathematical tool and not of fundamental significance.

The cosmological model considered here is not a realistic approximation of our universe, which appears to be extraordinarily isotropic on large scales. Aniso\-tropies are instead a local phenomenon that can be treated perturbatively. How this may be done has been suggested in section \ref{CommentsOnPertTheory} and full details will be given in a future paper \citep{RoserInPreparation}. Here the situation is more complicated since not only are the perturbations inhomogeneous but also must include matter terms and therefore further degrees of freedom. However, the finite-dimensional model developed here already suggests the possibility of a consistent quantum theory for the non-canonical Poisson brackets arising in the York-time Hamiltonian reduction.

\section*{Acknowledgements}

I am indebted to Antony Valentini for multiple discussions and a critical review of this manuscript.


\bibliographystyle{ieeetr}	
\bibliography{../../Bibloi}	



\end{document}